\documentclass[lettersize,journal]{IEEEtran}
\usepackage[utf8]{inputenc}
\usepackage{authblk}
\usepackage{setspace}
\usepackage{graphicx}
\graphicspath{ {./figures/} }
\usepackage{subcaption}
\usepackage{amsmath}
\usepackage {color}
\usepackage[table]{xcolor}
\usepackage{tabularx}
\usepackage{url}
\UseRawInputEncoding
\usepackage{amsmath}
\usepackage{lineno}
\usepackage{verbatim}

\usepackage[backend=bibtex,style=numeric,sortcites,natbib=true,sorting=none]{biblatex}

\title{Near-Space Communications: the Last Piece of 6G Space-Air-Ground-Sea Integrated Network Puzzle}

\author[1]{Hongshan Liu}
\author[1,2]{Tong Qin}
\author[2*,3,4,5]{Zhen Gao}
\author[2,3*]{Tianqi Mao}
\author[1]{Keke Ying}
\author[1]{Ziwei Wan}
\author[1]{Li Qiao}
\author[2,3]{Rui Na}
\author[2,3]{Zhongxiang Li}
\author[2,3]{Chun Hu}
\author[1]{Yikun Mei}
\author[2,3,4,5]{Tuan Li}
\author[6,7]{Guanghui Wen}
\author[2,3]{Lei Chen}
\author[8]{Zhonghuai Wu}
\author[9]{Ruiqi Liu}
\author[10]{Gaojie Chen}
\author[11*]{Shuo Wang}
\author[2,3]{Dezhi Zheng}

\affil[1]{School of Information and Electronics, Beijing Institute of Technology, Beijing 100081, China.}
\affil[2]{MIIT Key Laboratory of Complex-Field Intelligent Sensing, Beijing Institute of Technology, Beijing 100081, China.}
\affil[3]{Advanced Research Institute of Multidisciplinary Sciences 100081, Beijing Institute of Technology, Beijing, China.}
\affil[4]{Yangtze Delta Region Academy of Beijing Institute of Technology, Beijing Institute of Technology, Jiaxing 314000, China.}
\affil[5]{Advanced Research Institute of Multidisciplinary Science, Beijing Institute of Technology, Jinan 250307, China.}
\affil[6]{Department of Mathematics, Southeast University, Nanjing 210096, China.}
\affil[7]{Advanced Research Institute
	of Multidisciplinary Science, Beijing Institute of Technology, Beijing 100081,
	China.}
\affil[8]{Beijing Key Laboratory for Precision Optoelectronic Measurement Instrument and Technology, School of Optics and Photonics, Beijing Institute of Technology, Beijing 100081, China.}
\affil[9]{Wireless and Computing Research Institute, ZTE Corporation, Beijing 100029, China.}
\affil[10]{5GIC \& 6GIC,
	Institute for Communication Systems, University of Surrey, GU2 7XH Guildford, U.K.}
\affil[11]{China Academy of Space Technology, Beijing 100094, China.}
\affil[*]{Correspondence should be addressed to Zhen Gao: gaozhen16@bit.edu.cn; Tianqi Mao: maotq@bit.edu.cn; Shuo Wang: evan3210@sina.com}

\date{}

\begin{document}
	
	\maketitle
	
	\begin{abstract}
		
		This article presents a comprehensive study on the emerging near-space communications (NS-COM) within the context of space-air-ground-sea integrated network (SAGSIN). Specifically, we firstly explore the recent technical developments of NS-COM, followed by the discussions about motivations behind integrating NS-COM into SAGSIN. To further demonstrate the necessity of NS-COM, a comparative analysis between the NS-COM network and other counterparts in SAGSIN is conducted, covering aspects of deployment, coverage, channel characteristics and unique problems of NS-COM network. Afterwards, the technical aspects of NS-COM, including channel modeling, random access, channel estimation, array-based beam management and joint network optimization, are examined in detail.  Furthermore, we explore the potential applications of NS-COM, such as structural expansion in SAGSIN communication, civil aviation communication, remote and urgent communication, weather monitoring and carbon neutrality.  Finally, some promising research avenues are identified, including stratospheric satellite (StratoSat) -to-ground direct links for mobile terminals, reconfigurable multiple-input multiple-output (MIMO) and holographic MIMO,  federated learning in NS-COM networks, maritime communication,  electromagnetic spectrum sensing and adversarial game, integrated sensing and communications, StratoSat-based radar detection and imaging, NS-COM assisted enhanced global navigation system, NS-COM assisted intelligent unmanned system and free space optical (FSO) communication.  Overall, this paper highlights that the NS-COM plays an indispensable role in the SAGSIN puzzle, providing substantial performance and coverage enhancement to the traditional SAGSIN architecture.
	\end{abstract}
	
	
	\section{Introduction}
	The advent of the 5th generation (5G) technology has revolutionized the way we think about communication networks, enabling a plethora of new use cases and scenarios that were not possible with previous generations of wireless technology.
	5G offers significantly higher data rates compared to its predecessors, enabling faster downloads and smoother streaming experiences. In July 2020, the 3rd Generation Partnership Project (3GPP) successfully finalized the development of standard specifications for release 16, which is the first release of 5G. The 5G technology framework is anticipated to provide extensive support for a diverse array of prospective applications, classified into three principal usage scenarios: Enhanced Mobile Broadband (eMBB), Massive Machine Type Communications (mMTC) and Ultra Reliable Low Latency Communications (URLLC) \cite{ITU}. Moreover, 5G boosts the network capacity, accommodating a vast number of connected devices simultaneously. These attributes make 5G a promising advancement in wireless communication.
	
	However, 5G does come with its limitations. One of the primary challenges is its limited coverage in rural and remote areas, where the deployment of infrastructure can be economically and logistically challenging. Additionally, 5G's high-frequency signals have difficulty penetrating obstacles like buildings, resulting in possible coverage gaps especially in urban canyons where densely constructed tall buildings   increase multipath effects caused by signal reflections. 
	
	The limitations of existing networks, including capacity constraints, latency issues and energy inefficiencies, necessitate this next-generation leap. The 6th generation (6G) aims to create a hyper-connected, intelligent and immersive digital ecosystem \cite{10298069,8284057}. To meet these requirements, it needs to overcome the limitations of 5G by providing lower latency, seamless connectivity in diverse environments, as well as empower new frontiers in technology, enabling the efficient management of resources and services \cite{6genable}. 
	In response to these demands, 6G aims to provide a comprehensive solution, seamlessly integrating terrestrial, airborne and spaceborne networks, which is referred to as the space-air-ground-sea integrated network (SAGSIN). This integration will ensure uninterrupted communication across various environments, even in remote or other challenging locations. 
	
	\subsection{The Development of SAGSIN}
	
	The existing 5G communication networks predominantly rely on terrestrial communication infrastructure \cite{VT_Mag}, which, however, exhibits several notable limitations. For instance, as future high data transmission rates venture into much drastic frequency band, extremely high path loss will render traditional  hexagonal cellular networks inadequate in assuring signal quality within shadow fading regions. Thus, there is a growing need for finding signal transmission paths in a more flexible way. Furthermore, in the event of natural or man-made disasters, terrestrial base stations (BSs) in affected areas are susceptible to damage and their reconstruction can be exceptionally challenging, often hindering relief efforts. Additionally, to address issues of coverage and capacity, 5G necessitates an increased deployment of small cell BSs and network infrastructure, incurring substantial costs and complex deployment logistics. To mitigate these challenges, there is a growing imperative to extend communication from the tradition terrestrial network into SAGSIN, where expansions in vertical dimension becomes pivotal.
	
	It is expected that one of the key objectives of future 6G communication systems is the seamless integration of advanced multimedia services over heterogeneous networks \cite{10298069}.
	In the ever-evolving landscape of communications, the SAGSIN has emerged as a cutting-edge paradigm that promises to revolutionize how we connect and exchange information \cite{jiajialiu,9729746}. A traditional SAGSIN scenario mainly includes three segments: spaceborne network, airborne network and terrestrial network, each can work independently or inter-operationally. By integrating heterogeneous networks among these segments, a highly efficient hierarchical communication network is built, providing a deep integration of systems, technologies and applications.
	
	The spaceborne network \cite{LEO_Mag, joint} includes the geostationary Earth orbit (GEO), medium Earth orbit (MEO) and low Earth orbit (LEO) satellites that operate at different altitudes. The airborne network \cite{parcel, VT_Mag, Future_Aerial} mainly consists of airships, balloons, aircraft and unmanned aerial vehicles (UAVs). The terrestrial network contains ground-based infrastructures and user devices in urban areas as well as areas that lack infrastructure, such as rural areas, ocean areas, disaster areas and dense urban areas. To maintain sufficient capacity, energy efficiency and reliability, the mentioned areas are required to be interconnected to create a comprehensive and robust communication ecosystem, in order to achieve ubiquitous coverage on a global scale. 
	{\color{black} For example, an uplink NOMA scenario was examined in \cite{add1} within a satellite-aerial-ground integrated network, where multiple users engaged in communication with a satellite aided by a UAV serving as an aerial relay.}
	

	\subsection{Motivations of Introducing Near-Space Communication Network}
	
	\begin{figure*}[h]
		\centering
		\includegraphics[width=0.8\textwidth]{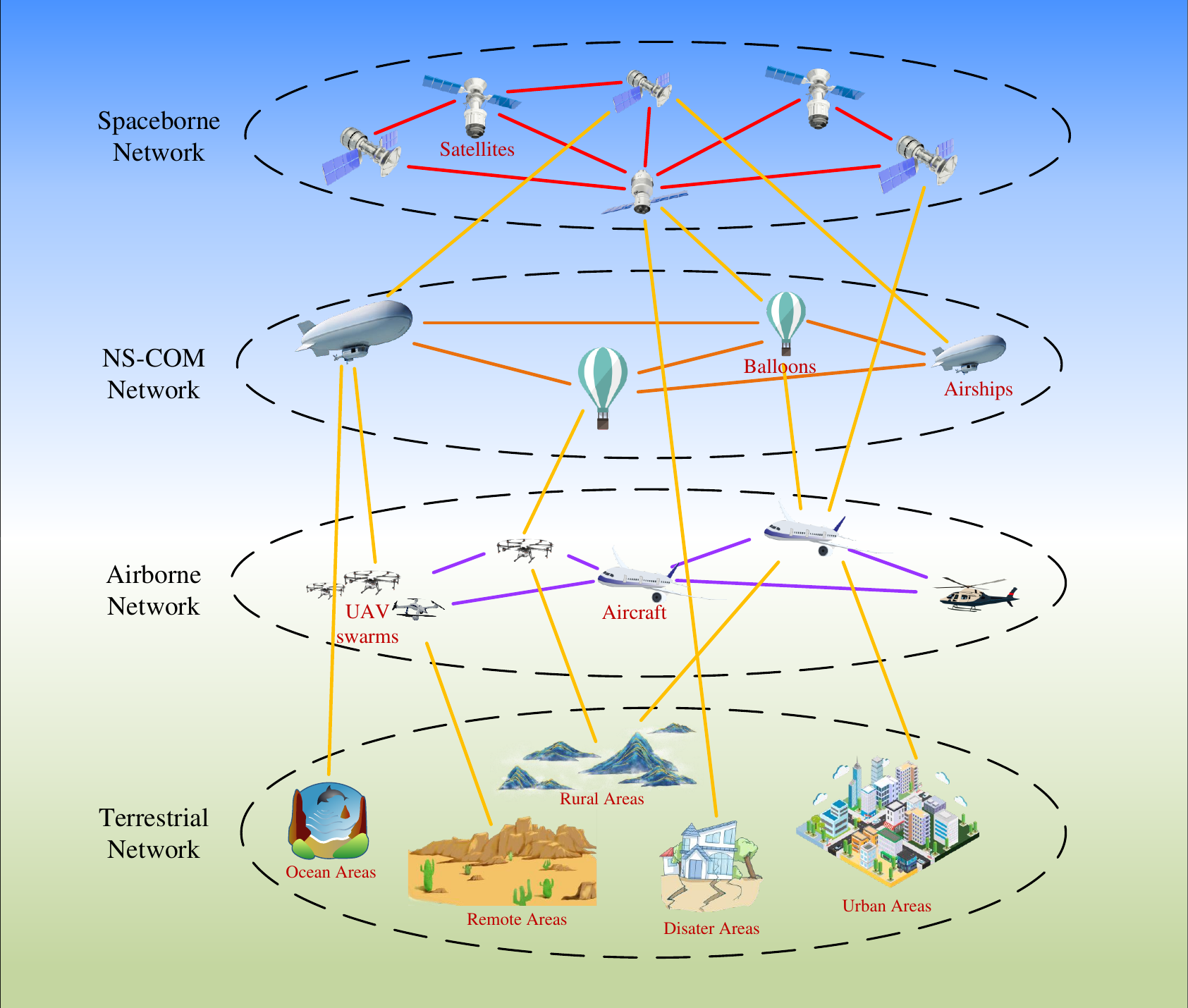}
		\caption{Schematic diagram of the SAGSIN architecture, consisting of  spaceborne, near-space, airborne and terrestrial networks.}
		\label{fig:1}
	\end{figure*}
	
	The classical SAGSIN architecture have shown remarkable capabilities in providing massive connectivity by integrating spaceborne, airborne and terrestrial cellular networks, which, however, encounters inherent technical challenges against their respective practical implementation.
	Satellite systems, while being able to offer global coverage in theory, lead to significant latency and cost issues for real-time applications due to their extremely high altitude. 
	Small airborne platforms such as UAVs can offer several benefits like rapid deployment, flexible maneuverability and the ability to provide temporary coverage \cite{UAV}. However, its limited coverage makes it difficult to provide connectivity to vast area. Additionally, they also face challenges including limited endurance, regulatory constraints in shared airspace and vulnerability to adverse weather conditions. 
	Moreover, the terrestrial network, struggles to extend coverage to remote area economically and are susceptible to service disruptions during disasters. It may also experience congestion and reduced quality of service with the unprecedented increase of mobile users in the 6G era. 
	
	On the other hand, the philosophy of near-space communications (NS-COM) is gradually emerging. Near-space locates between 20 km and 100 km above the Earth's surface, characterized by its reduced turbulence intensity, extremely low air pressure down to 0.05 times the standard atmospheric pressure and  temperatures as low as minus 60 degrees Celsius.
	These unique physical characteristics can guarantee airships and high-altitude balloons in this region to maintain near-geostationary positions in the stratosphere for extended periods, often spanning several weeks or months, even with heavier payloads. The near-space airships and balloons have collectively been referred as stratospheric satellites  (StratoSats) , which compose the NS-COM network, an independent networks entity separate from the traditional airborne network. StratoSats are also called  high altitude platform stations (HAPSs) or near-space platform stations (NSPSs) \cite{Xiao}. They are able to carry heavy payloads for extended periods and offer controlled movement, providing a stable platform for instruments as well as offering comparatively higher energy efficiency. For example, balloons follow passive drift patterns due to stratospheric winds, making them ideal for passive observation and data collection for the purpose of weather forecast \cite{9941044}. 
	Aside from the aforementioned merits, the distinctive location of near-space allows the NS-COM network to address the shortcomings of traditional spaceborne, terrestrial and airborne networks. Positioned at the core of the network, NS-COM plays a pivotal role in facilitating seamless cross-network operations and resource allocation within the SAGSIN framework, which is exhibited in Fig. \ref{fig:1}. StratoSats can serve as both a supplementary component and an independent service provider. As a supplementary part of SAGSIN, StratoSats act as crucial bridges, realizing the seamless integration between different networking layers. Specifically, their strategic position between the lower atmosphere and space makes them qualified for efficient cross-layer communication relaying.
	Additionally, as independent service providers, StratoSats offer several advantages over other networks \cite{1025040}. Their extended endurance allows for continuous and persistent coverage over vast geographical areas, overcoming the limitations of airborne platforms with their limited operational duration and reducing the long transmission delay of traditional satellites. Operating in the stratosphere, above regulated airspace, StratoSats simplify deployment and management, reducing the complexities associated with terrestrial and airborne network \cite{9900369}.

	
	NS-COM present a promising addition to SAGSIN by addressing the drawbacks of existing communication networks, making them the final piece of the SAGSIN puzzle. It can be utilized in various scenarios, such as navigation, positioning systems, remote sensing and emergency communications, making them a promising solution for the Internet of Everything era. They harness the best features of the existing  communication systems, providing high-bandwidth, low-latency connectivity over a wide area without the need for extensive ground infrastructures. 
	{\color{black}
		The industry has shown significant interest in the design and production of StratoSats, leading to the development of numerous projects and products over the past decade. Among them, those that are relatively representative include Project Loon, Stratobus and Yuanmeng.
		Project Loon employs high-altitude balloons positioned in the stratosphere at altitudes ranging from 18 km to 25 km to establish an aerial wireless network capable of providing internet access to remote areas. These balloons utilize patch antennas, which are directional, to transmit signals. Initially, its communication utilizes the  unlicensed 2.4 G Hz and 5.8 G Hz bands, providing speeds comparable to Third Generation Mobile Telecommunications Technology (3G). However, the system later transitioned to Long-Term Evolution (LTE), utilizing the cellular spectrum in collaboration with local telecommunication operators. The typical lifespan of the balloons is around 100 days \cite{loon}.
		Stratobus, with a length of 140 meters, operates autonomously within the stratosphere at an altitude of 20 kilometers. It possesses the capability to cover a terrestrial horizon spanning up to 500 kilometers, rendering it particularly apt for environmental monitoring missions. In terms of observational tasks, Stratobus is equipped with both radar and optical imaging payloads, thereby enabling continuous surveillance irrespective of weather conditions. Its potential extends to the telecommunications sector. Weighing nearly seven metric tons and featuring dimensions of 115 meters in length and 34 meters in diameter at its broadest point, Stratobus accommodates payloads weighing up to 450 kilograms and boasts an 8 kW power rating. Positioned within the zone between the tropics, characterized by winds below 90 km/h, it holds promise for providing uninterrupted service year-round on a 24/7 basis \cite{stratobus}.
		Airship Yuanmeng represents China's inaugural near-space platform designed for both military and civilian applications. With a payload capacity of 300 kg and a volumetric capacity of approximately 18,000 cubic meters, Yuanmeng is equipped with advanced systems for wideband communication, relay, high-definition observation, and spatial imaging. The platform boasts an impressive mission duration of up to 6 months. Inaugurating its first flight in 2015, Yuanmeng underwent a 48-hour test flight that ascended to a near-space altitude of 20 kilometers to assess its operational systems and capabilities \cite{ym}. 
	}
	
	\begin{figure*}[h]
		\centering
		\includegraphics[width=0.8\textwidth]{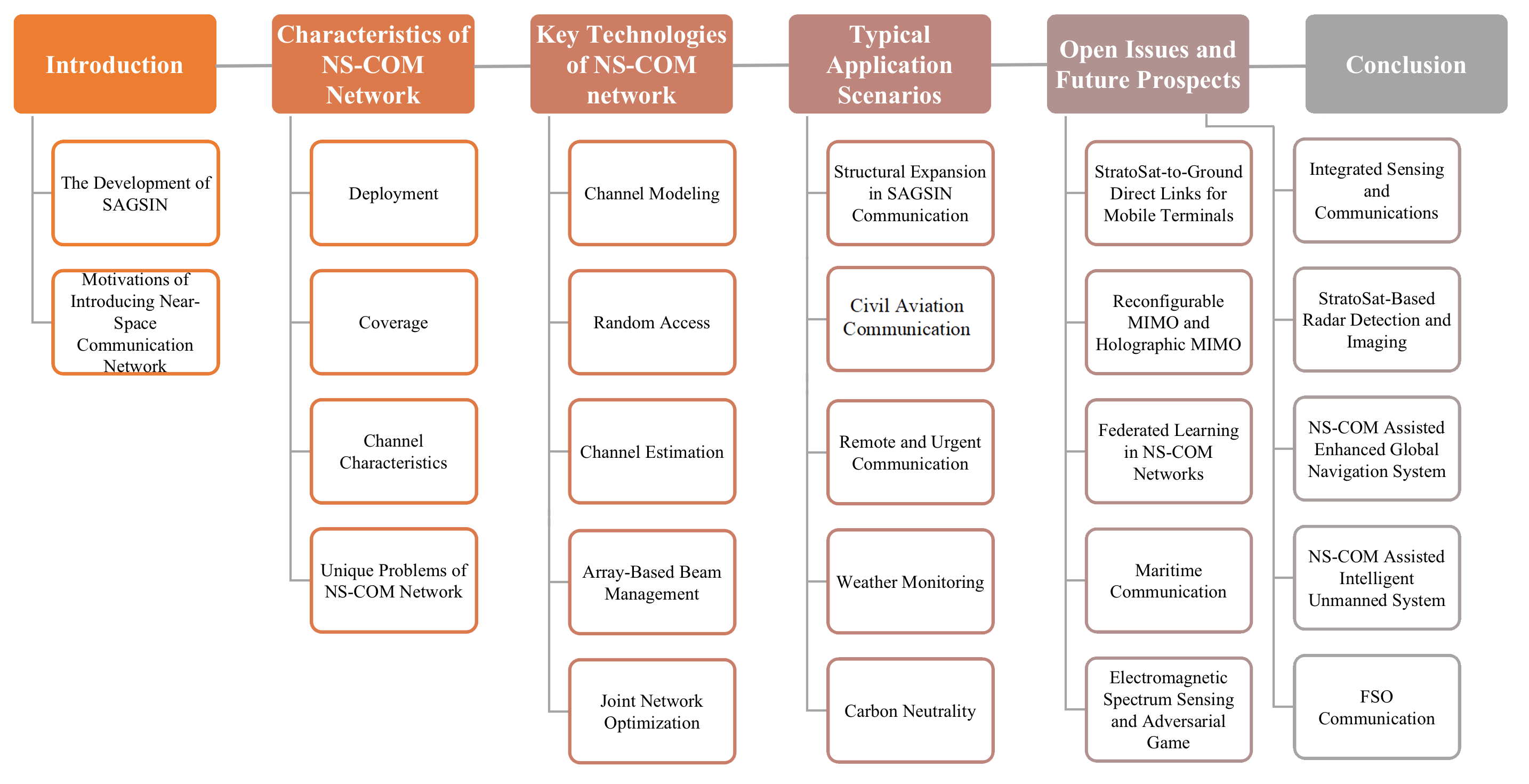}
		\caption{\color{black}The outline of this paper.}
		\label{fig:0}
	\end{figure*}
	\color{black}{
		To provide researchers with a more comprehensive understanding of NS-COM and inspire further advances, the rest of the paper will present a detailed review of NS-COM. As illustrated in Fig. \ref{fig:0}, this survey is organized as follows.}
	\color{black}
	In Section 2, we particularly introduce the characteristics of NS-COM network while comparing it with the other networks in SAGSIN. 
	Section 3 discusses several technical features of NS-COM.
	Section 4 highlights some typical application scenarios for NS-COM.
	In Section 5, we explore candidate techniques for NS-COM, as well as discuss some open issues therein.
	Finally, the conclusions are summarized in Section 6.
	
	Compared with existing relevant survey papers, this article concentrates more on how NS-COM network expands coverage and enhance quality of service (QoS) in SAGSIN, ensuring better service for 6G and beyond communication networks. Relevant advanced technologies, including channel modeling, random access, channel estimation, array-based beam management and joint network optimization are discussed. We also focus on the benefits and challenges of several potential cutting-edge technologies such as StratoSat-to-ground direct links for mobile terminals  to support the implementation of NS-COM for 6G and beyond communications in the future.
	
	\section{Characteristics of NS-COM Network}
	
	To demonstrate the indispensability of NS-COM, we will provide a comprehensive investigation on its characteristics in terms of deployment, coverage and channel characteristics, in comparison with the traditional terrestrial/airborne/spaceborne networks.
	
	\subsection{Deployment}
	Terrestrial network has been extensively established in urban and populated areas, providing reliable connectivity for everyday communication needs. However, its post-deployment flexibility is significantly inferior compared to the other networks. The airborne network, represented by UAVs, offers rapid deployment capabilities, making them suitable for temporary or emergency coverage in remote or disaster-affected regions. However, the flexibility comes with drawback of low stability.
	For space communications, satellite nodes are typically positioned in specific orbits to provide global coverage, which, however, suffers from high maintenance costs and launch costs.
	StratoSats, positioned in the stratosphere above commercial aircraft but below traditional satellites, boost an advantageous mid-altitude deployment. By combining the durability of terrestrial network and the versatility of spaceborne network, it compensates for their respective shortcomings. A comparative summary regarding different aspects of deployment is presented in Table \ref{tab:deployment-comparison}.
	\begin{table*}[ht]
		\centering
		\caption{Comparison of Deployment Features.}
		\label{tab:deployment-comparison}
		\begin{tabularx}{\textwidth}{|X|X|X|X|X|X|}
			\hline
			\rowcolor{yellow}
			\textbf{}            & \textbf{Terrestrial Network}          & \textbf{Airborne Network}          & \textbf{NS-COM Network} & \textbf{Spaceborne Network}   \\ \hline
			\textbf{Deployment Method}         & Laying cables, erecting cell towers & UAV launch and aerial deployment  & High-altitude balloon or aircraft launch & Rocket launch into orbit          \\ \hline
			\textbf{Deployment Cost}           & High initial investment for infrastructure & Moderate to high for UAVs and HAPs  & Moderate for stratospheric platforms & High for satellite launch         \\ \hline
			
			\textbf{Maintenance Mode}      & Regular inspections and repairs          & Frequent UAV replacements and maintenance & Less frequent & Limited due to distant orbits    \\ \hline
			\textbf{Loadability}              & No loadability constraint       & Moderate payload capacity of UAVs  & Substantial payload capacity for instruments & Varies with satellite design      \\ \hline
			\textbf{Power Consumption}        & Low power for ground infrastructure      & UAVs have moderate power consumption & Moderate power usage for stratospheric platforms & Varies with satellite mission     \\ \hline
		\end{tabularx}
	\end{table*}

\textbullet { \textbf{Deployment Method}}: 
Terrestrial network relies heavily on physical infrastructure, involving the laying of cables and erecting cell towers on the ground. While these networks provide stable and reliable coverage in densely populated areas, they face limitations in reaching remote or challenging terrains, where deployment might be impractical or economically unviable.
Airborne network leverages the use of UAVs for deployment. UAVs offer rapid mobility and flexibility, allowing them to establish temporary connectivity in disaster-stricken or remote regions, albeit with limitations in flight endurance and regulatory considerations.
For spaceborne network, traditional satellites, launched into GEO, MEO, or LEO orbits, deliver global coverage but demand complex and costly rocket deployments. 

StratoSats stand out as a unique deployment approach by reaching the stratosphere. StratoSats can be deployed using various methods such as lifted by high altitude balloons or launched by specialized  aircrafts. This flexibility allows for cost-effective and more frequent deployment opportunities, enabling quicker response times for mission deployment or network reinforcement, bringing communication services to places that traditional terrestrial and even some satellite networks struggle to reach.

\textbullet { \textbf{Deployment Cost}}: For terrestrial network, initial costs involve infrastructure investment, including cables and potentially cell towers. However, operational expenses can be relatively lower, making them cost-effective for widespread coverage in populated areas.
Airborne network, utilizing UAVs for deployment, entails moderate to high costs. The development, acquisition and operation of UAVs add to the deployment expenses. The need for frequent replacements and maintenance of UAVs due to flight endurance limitations can also result in ongoing operational expenses, impacting the overall cost-effectiveness of airborne network.
Traditional satellite deployment represents the highest initial cost among the layers. The complex and costly rocket launches required to place satellites into their respective orbits contribute significantly to the deployment expenses. Moreover, the maintenance and operational costs associated with traditional satellites can be substantial due to the need for monitoring, ground station infrastructure and satellite servicing.

StratoSats offer an economical approach to deployment costs, striking a balance that is both sustainable and cost-effective. While they do require a moderate investment for launching high-altitude balloons or specialized aircraft, their extended endurance and ability to cover vast geographical areas demonstrate their inherent value over the long term. This balance is crucial in a world where the demands for continuous communication are ever-increasing. StratoSats, by combining affordability with efficiency, stand as a viable solution for a wide range of applications, from scientific research and disaster management to communication relays and Earth observation. 

\textbullet { \textbf{Maintenance Mode}}:
Terrestrial network requires regular inspections and repairs of physical infrastructure, such as cables and cell towers, to ensure uninterrupted connectivity. Although terrestrial network offers cost-effective maintenance once established, their reliance on physical components demands continuous monitoring and swift response to potential issues.
Airborne network, utilizing UAVs and other aerial platforms, presents unique maintenance challenges. The maintenance mode involves managing UAV fleets, conducting regular inspections and promptly addressing any malfunctions to ensure continuous service delivery. 
Maintaining spaceborne network operations poses several challenges. Due to the harsh space environment with extreme cold temperatures and high radiation, equipments face severe degradation, however the repairability in space is limited. 

By contrast, StratoSats platforms can stay aloft for extended periods, reducing the need for frequent maintenance. StratoSats possess the unique capability to maintain their high-altitude positions for significantly extended duration, which inherently diminishes the necessity for regular maintenance. This characteristic of StratoSats not only promotes cost-efficiency but also underscores their advantage in terms of operational longevity. Additionally, StratoSats benefit from reduced regulatory burdens as they operate in near-space, away from heavily regulated airspace.

\textbullet { \textbf{Loadability}}:
The loadability problem across different layers in the SAGSIN demands careful consideration of the network's capabilities and scalability to meet increasing data demands. Terrestrial network, being deployed on the ground, has no inherent loadability restriction and can handle substantial data traffic in densely populated areas.
Airborne network utilizing UAVs offer moderate loadability capabilities whereas limited payload capacity \cite{9526159}, indicating restricted device capacity for transmission, processing or accommodation. Typical commercial drones used in airborne network can carry payloads ranging from a few kilograms to around 20 kg, limiting their capacity to handle large-scale data traffic.
Traditional satellites, positioned in various orbits, exhibit diverse loadability capacities based on their respective missions and designs. Typical loadability values for GEO satellites range from several hundred kilograms to a few tons, for MEO satellites it varies from hundreds of kilograms to around one ton. And for LEO satellites, it ranges from a few kilograms to several hundred kilograms.

With substantial payload capacity up to almost 500 kg of for various communication instruments \cite{elevate}, StratoSats can provide extensive coverage and efficiently handle data traffic over large areas. Simultaneously, this payload facilitates the provision of multiple types of services, for example, tasks such as collecting extensive measurement data within a typhoon, which requires the deployment of a substantial number of communication and sensing nodes, are well within the capabilities of StratoSats.

\textbullet { \textbf{Power Consumption}}:
The power consumption of different communication networks in the SAGSIN is a critical factor influencing their operational efficiency and environmental impact. Terrestrial network generally has lower power consumption but can experience an increase in high data demand areas. Airborne network, such as UAVs, have higher power consumption due to propulsion and electronics requirements. Traditional satellites require energy on the propulsion systems for orbit adjustments and station-keeping maneuvers. The energy demands also include thermal control systems to regulate the satellite's temperature, ensuring proper functioning of electronic components. 

StratoSats offer a balanced power consumption profile, relying on solar energy, making them suitable for long-duration missions. Their ability to remain aloft for prolonged periods,  enabling persistent and continuous coverage over large geographic areas, allowing for seamless transitions in areas requiring sustained connectivity. 
The prolonged and continuous operation of StratoSats to support extended missions lasting for days or even weeks necessitates a robust and sustainable energy supply.
As a result, StratoSats predominantly rely on solar panels as their primary energy source to harness power from the sun. The energy extraction is supplemented by the integration of hybrid energy sources, including light gas (e.g., hydrogen and helium), wind and radio frequency (RF) waves to ensure reliable energy generation even in variable weather conditions.
At the altitudes where StratoSats operate, the wind in the stratosphere is generally weaker compared to those in the troposphere. However, this relatively modest airflow doesn't deter StratoSats from optimizing their flight by adjusting their orientation to harness wind power more efficiently. This strategic adjustment allows them to propel forward with reduced propulsion power consumption.

\subsection{Coverage}
In this part, we will delve into the  coverage differences of each network layer. Specifically, we discuss the coverage ability of different layers from the aspects of time and space. 
In the time dimension, the focus is on ensuring continuity of service, where communication networks must maintain uninterrupted connectivity under various circumstances. Factors such as network reliability, resilience to disruptions and adaptability to changing demands play critical roles in addressing this aspect of the coverage problem.
Conversely, the space dimension revolves around the coverage range, which entails how extensively a network can reach different geographic areas. Networks should be capable of offering localized coverage for specific regions, extended reach to remote or rural locations and even global coverage for broader applications.
\subsubsection{Temporal Coverage}
Terrestrial network offers good continuity of service as they can operate continuously once established. The network relies on fixed ground-based infrastructure, disruptions may occur due to natural disasters or physical obstacles. However, quick restoration measures and redundant systems ensure timely recovery and continuous connectivity to users.

Airborne network provides flexible and mobile solutions for communication. However, they face limitations in terms of continuity of service due to power constraints. These platforms may serve users for limited periods before requiring recharging or redeployment, making them suitable for temporary or emergency coverage.

The NS-COM ensure robust connectivity and are less susceptible to disruptions caused by adverse weather conditions or ground-based obstacles, which makes it more time-efficient to emergency situations in comparison with terrestrial network. Although StratoSats may not be as  flexible and convenient as airplanes and UAVs due to its bulky appearance, the larger volume offers longer endurance as a compromise. Since StratoSat operates relatively stationary with respect to the Earth, small cell switching is avoided, which simplifies the scheduling tasks. 

For spaceborne network, LEO satellites offer extensive coverage and can form large, densely deployed constellations \cite{9312798}. However, they may suffer from intermittent interruptions due to their orbital characteristics. MEO and GEO satellites, while providing broader coverage, may be vulnerable to weather and atmospheric influences, affecting their capabilities to provide continuous service.

\subsubsection{Spatial Coverage}
Spatial coverage is a critical aspect of the SAGSIN that determines the geographic reach and accessibility of communication services. The spatial coverage capabilities of different layers in the SAGSIN are diverse and complementary.

Terrestrial network excels in providing localized coverage in urban and densely populated areas. It deploys infrastructure such as cell towers and BSs to create communication cells, ensuring reliable connectivity within limited geographic regions. The typical coverage range of a terrestrial cell is around hundreds of meters in diameter. However, the coverage area is inherently limited by the physical reach of terrestrial infrastructure, making it challenging to extend connectivity to remote and sparsely populated areas.

Airborne network, particularly UAV-based solutions and high-altitude platforms, offers a unique advantage in extending coverage to remote and hard-to-reach regions \cite{9380358}. Their mobility allows them to traverse vast areas, making them suitable for temporary deployments in disaster-hit or emergency scenarios. The typical operating height for UAV-based network ranges from a few hundred meters to several kilometers, enabling them to cover areas with a radius of hundreds of meters during each flight. However, their coverage area remains limited compared to stratospheric and traditional satellites.

NS-COM provides extended coverage over larger geographic regions compared to airborne network. Positioned in the stratosphere, they can cover extensive areas, making them suitable for providing connectivity to rural and remote regions without the need for complex ground infrastructure. StratoSats can cover areas with a radius of up to 500 km \cite{stratobus}, depending on their positioning and altitude. It's ability of providing an enormous coverage area is one of its most significant advantages compared to terrestrial network and airborne network, making it an ideal solution for remote and rural areas \cite{9773096}.

Traditional satellites operate in different orbits, providing varying coverage ranges. LEO satellites orbit close to the Earth at altitudes of around 500 to 2,000 km and can form large constellations to achieve global coverage \cite{9312798}. They typically cover areas with a radius of hundreds of kilometers per satellite pass.

\subsection{Channel Characteristics}

Table. \ref{tab:channel_characteristics} presents the comparison of channel characteristics among different layers in the SAGSIN. 
Terrestrial and airborne network primarily operates in Sub-6 Gigahertz (GHz) bands, benefiting from moderate path loss and low transmission delay. StratoSats, on the other hand, utilize millimeter wave (mmWave) frequencies to achieve high data rates, while traditional satellites use multiple bands. 
In terrestrial and near-space network, interference is generally low, ensuring reliable communication. However, it is essential to consider environmental conditions, such as weather and attenuation, which may impact airborne and stratospheric layers. Understanding these characteristics is vital for designing efficient communication systems that can perform optimally in diverse environments.

	\begin{table*}[htbp]
		\centering
		\caption{Channel Characteristics of Different Layers in SAGSIN.}
		\label{tab:channel_characteristics}
			\begin{tabularx}{\textwidth}{|>{\centering\bfseries}X|X|X|X|X|}
				\hline
				\rowcolor{yellow} \textbf{Aspect} & \textbf{Terrestrial Network} & \textbf{Airborne Network} & \textbf{NS-COM Network} & \textbf{Spaceborne Network (LEO, MEO, GEO)} \\
				\hline
				\textbf{Frequency Band} & Sub-6 GHz (e.g., 2.4 GHz, 3.5 GHz) & mmWave and Sub-6 GHz (e.g., 2.4 GHz, 5.8 GHz) & mmWave (e.g., 47.2-47.5 GHz) and Sub-6 GHz (e.g., 2.7 GHz) & Multiple Bands (e.g., C-band, Ku-band, Ka-band) \\
				\hline
				\textbf{Path Loss} & Moderate & Low to Moderate & Moderate & High \\
				\hline
				\textbf{Transmission Latency }& Very low & Low & Low to moderate & Moderate to high\\
				\hline
				\textbf{Doppler Effect} & N/A & Very High & Negligible & High for LEO/MEO, negligible for GEO\\
				\hline
				\textbf{Environmental Condition} & N/A & Susceptible to weather & Quasi-static  & Affected by ionospheric fluctuations\\
				\hline
			\end{tabularx}%
		\end{table*}


	\textbullet { \textbf{Frequency Band}}: 
	The frequency bands used in different networks vary based on their unique characteristics and communication requirements. In terrestrial network, various frequency bands have been utilized across different generations of cellular technology. The latest 5G network has extended to higher frequency ranges, such as the sub-6 GHz bands (e.g., 3.5 GHz) and mmWave bands (e.g., 28 GHz), to unlock unprecedented data speeds and low latency.
	Airborne network, specifically UAVs, often adopt the sub-6 GHz frequency band (e.g., 2.4 GHz and 5.8 GHz). These bands offer a good balance between data rates, transmission range and mobility support for UAVs. Additionally, the sub-6 GHz bands are widely available and globally harmonized, facilitating seamless communication during UAV flights.
	Traditional satellites, comprising GEO, MEO and LEO satellites, deploy various frequency bands to optimize their performance. GEO satellites mainly use the C-band (4 to 8 GHz) and Ku-band (12 to 18 GHz) for broadcasting and data services. MEO satellites leverage the Ka-band (26.5 to 40 GHz) for broadband internet services. LEO satellites mainly utilize UHF-band (0.3 to 1 GHz), L-band (1 to 2 GHz) and Ku-band frequencies for low-power communication with ground terminals. Notably, SpaceX's Starlink constellation employs a mix of these frequency bands to offer global internet connectivity.
	
	StratoSats, operating at high altitudes in the stratosphere, typically utilize the mmWave frequency band. At the 1997 World Radiocommunication Conference (WRC-97), the frequency bands, specifically 47.2-47.5 GHz (downlink) and 47.9-48.2 GHz (uplink), were initially designated for global use by StratoSats \cite{ar}. The mmWave spectrum allows for high-speed data transmission, extensive bandwidth and precise beamforming capabilities \cite{8761640}. However, due to significant rain fading challenges and the limited maturity of technical solutions in the higher frequency bands, during the World Radiocommunication Conference in 2019 (WRC-19), it was decided to allocate certain frequency bands below 2.7 GHz for utilization by StratoSats \cite{ar}. This adjustment aims to address the mentioned technical limitations and enhance the effectiveness of StratoSat communication.

	For most air-to-air networks (including spaceborne network, NS-COM network and airborne network), short-haul communication can also be accomplished using terahertz (THz) and Free-Space Optical (FSO) communication \cite{6882305,7321055,7490372}. Especially for communication with vehicles above the stratosphere, where atmospheric molecule absorption and rain attenuation can be neglected \cite{8656538,8901159}.
	
	\textbullet { \textbf{Fading}}:
	Since the transmission distance issue will be further discussed in the transmission latency section, large scale fading is ignored in the following discussion. Terrestrial network encounters significant shadow fading in urban settings due to obstructions like buildings and vegetation, resulting in signal attenuation. Fluctuations in signal strength occur as users move within the coverage area.
	Airborne network benefits from elevated positions, reducing shadow fading and enabling direct line-of-sight (LoS) communication. This advantage enhances signal propagation and coverage, especially in remote areas.
	Due to the near-vacuum conditions prevailing in space where satellites operate, the impact of signal attenuation is significantly minimized. Similar to StratoSats, since the transmission is close to free space propagation, the fading effect is almost neglectable.
	
	The NS-COM network experiences relatively lower fading compared to ground-based systems, as the StratoSats are positioned above atmospheric interference. Their higher elevation allows for extended communication range and improved signal transmission. The absence of scattering bodies within this environment further reduces signal attenuation, allowing for minimal interference or degradation in signal quality. Additionally, the flexible positioning of the StratoSat enables it to avoid unfavorable weather conditions, thereby enhancing the quality of connections.
	
	\textbullet { \textbf{Transmission Latency}}: 
	Transmission latency, a crucial factor in communication systems, varies significantly among satellite network. Terrestrial and airborne networks generally exhibit lower latency due to their shorter distance to users. However, traditional satellite network, including LEO, MEO and GEO satellites, encounters varying levels of latency.
	LEO satellites, positioned closer to the Earth's surface, offer low-latency communication, typically ranging from 1.3 milliseconds (ms) to 6.7 ms, while GEO satellites, positioned much farther away, exhibit higher latency, with typical one-way latency for up to almost 1200 ms.
	
	In comparison, StratoSat's shorter distance to the ground results in lower latency and path loss, with a signal delay of only 0.06 ms, making it more advantageous in operations such as beamforming and random access that require multiple uplink and downlink communications.
	
	\textbullet { \textbf{Doppler Effect}}:
	
	
	Doppler shift effects in terrestrial network can range from a few Hz to a few hundred Hz, depending on the speed of mobile users or vehicles. For example, with  highway speeds of 120 km/h, the Doppler shift can be in the range of tens of Hz.
	Typical velocities of UAVs during surveillance and reconnaissance missions range from 30 to 100 km/h, resulting in Doppler shifts in the range of a few hundred Hz to a few kHz. 
	LEO satellites, with velocities of up to 28,000 km/h and extremely high frequency band, encounter Doppler shifts in the range of a few kHz to tens of kHz, depending on their relative motion.
	MEO satellites experience lower doppler shifts, typically in the range of a few hundred Hz to a few kHz. 
	Since GEO satellites is quasi stationary to the Earth, it causes minimal doppler effect.
	
	StratoSats, being relatively stationary in the stratosphere at altitudes of around 20 km, experience minimal doppler shift or drift effects. Although in scenarios where users exhibit high mobility, the StratoSat platform performs at a much higher altitude than the moving platforms of users, therefore, it can be approximated that the StratoSat operates directly above the user. The relative speed between the user and the platform primarily occurs in the vertical direction of the line connecting the user and the platform, resulting in almost zero relative speed in the direction of the line, thereby causing minimal Doppler effect.

	\textbullet { \textbf{Environmental Conditions}}: 
	Terrestrial network is vulnerable to various weather conditions, including rain, snow and fog. Precipitation, especially heavy rain, can cause signal attenuation and increase path loss, leading to reduced coverage and signal quality. Fog and other atmospheric conditions can also scatter and absorb radio waves, affecting signal strength and causing signal fluctuations. Additionally, extreme weather events such as storms or hurricanes can cause damage to infrastructures, leading to service disruptions.
	Airborne network operates in the troposphere, where they may encounter atmospheric turbulence and changes in temperature and pressure. These conditions can cause signal fading and fluctuations, especially in non-LoS scenarios. UAVs operating at lower altitudes may be more affected by weather conditions compared to StratoSats at higher altitudes in the stratosphere.Traditional satellites, especially those in LEO, MEO and GEO orbits, can be significantly affected by atmosphere. Ionospheric fluctuations in the atmosphere can introduce signal scintillation and interference, impacting the reliability of traditional satellite systems. These factors underscore the advantages of StratoSats in providing stable and efficient communication services, especially in challenging environmental conditions.
	
	StratoSats operate above most weather systems in the troposphere. As a result, they experience relatively stable and favorable weather conditions, with minimal impact on signal propagation.  Additionally, StratoSat is highly resistant to interference due to the extremely high spatial sparsity in near-space, making it suitable for use in areas with a high level of electromagnetic interference, such as urban areas. 
	The atmospheric channel for NS-COM in the stratospheric layer exhibits lower attenuation, making it a suitable candidate for communication in higher frequencies, up to the THz band, which can provide up to tens of GHz ultra-broadband and terabit per second ultra-high peak data rates  \cite{8387211}. Additionally, the weak wind speed in the stratosphere results in low power consumption to maintain stability and beam accuracy for StratoSats.
	\subsection{Unique Problems of NS-COM Network}
	\color{black}
	\begin{figure*}[t]
		\centering
		\includegraphics[width=0.8\textwidth]{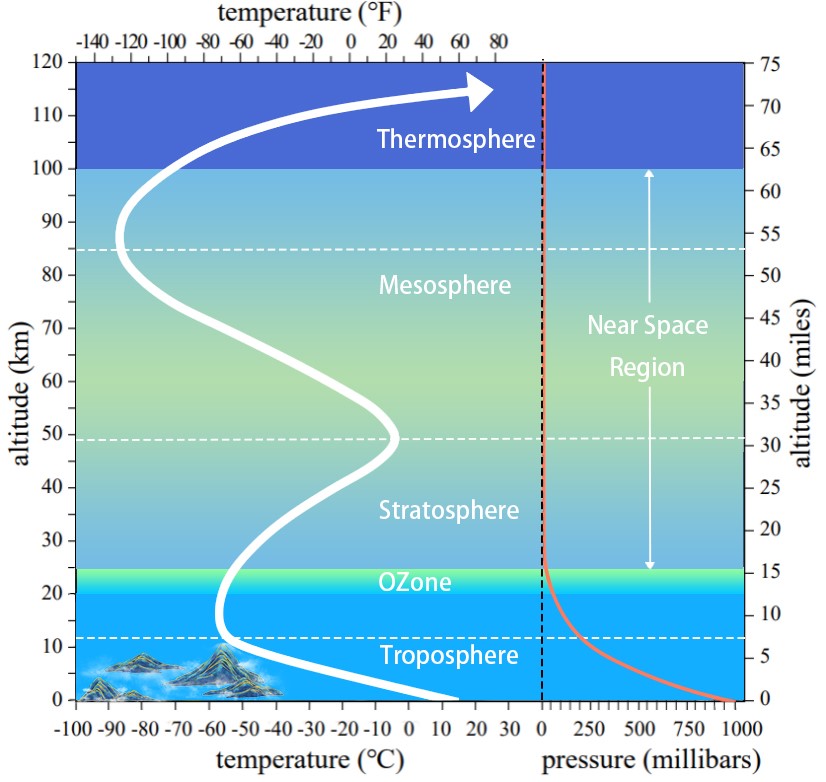}
		\caption{\color{black}The layers of Earth's atmosphere, where the near-space region is specified, with a white line showing the atmospheric temperature and a red line showing the pressure at various heights \cite{engg,Xiao}.}
		\label{fig:22}
	\end{figure*}
	NS-COM possesses distinct characteristics that render existing technologies used in spaceborne, airborne, and terrestrial networks unsuitable for near-space environments. The unique properties of NS-COM can be manifested in terms of channel conditions and beam reliability of the platforms.
	
	Although multipath effects almost barely exist in near-space, the atmospheric environment undergoes complex variations with altitude. The near-space environment of NS-COM networks exhibits environmental peculiarities compared to other networks, including high radiation, low temperature, low moisture and low atmospheric pressure, resulting in a propagation environment markedly different from those at other altitudes. The composition of the atmosphere above the Earth's surface, ranging from 20 km to 100 km, varies significantly with altitude, leading to large-scale changes in air pressure and temperature.
	Specifically, as illustrated in Fig. \ref{fig:22}, the near-space region encompasses the majority of the stratosphere, the mesosphere, and part of the thermosphere, where the air composition, temperature, and atmospheric pressure differ substantially. The near-space atmosphere displays varying constituents across different layers \cite{engg,Xiao}.  The temperature is generally lower in near-space but approaches zero at the top of the stratosphere due to ozone's radiation absorption. Temperature and density variations also result in non-uniform atmospheric pressure, posing challenges to conventional channel models for airborne and satellite networks. Refraction effects, exacerbated by the non-homogeneity of the transmission medium and aerosol scattering, particularly affect ultra long distance near-space-ground data links by causing non-negligible deviations of the beam directions \cite{Xiao}. Moreover, while atmospheric pressure is extremely low in most near-space regions, variations in atmospheric pressure near the bottom of the stratosphere are significant and cannot be ignored for channel modeling of StratoSat-aircraft links. The density of water vapor in near-space also induces additional frequency selectivity to the propagation channel and propagation loss. All these factors that increase the complexity of the channel render traditional channel modeling impractical as a result.
	
	Terrestrial or aerial network typically operates within a range of only a few hundred meters, lacking the significant variations in atmospheric conditions seen in near-space. Inter-satellite network experiences minimal attenuation, leading to simpler channel conditions. Microwave frequencies, especially the L-band, are commonly utilized for satellite-ground communication \cite{417777}, yet supporting high-rate services is unrealistic due to bandwidth limitations. The heavily congestion of the microwave frequency band also makes it challenging to allocate new frequency bands.

	StratoSats can use large-scale phased antenna arrays at mmWave or higher frequencies for highly directional data transmission. This approach meets payload budget constraints because of the compact size of mmWave-band antenna elements. However, narrow beamwidth requires precise beam alignment, leading to severe fading due to antenna misalignment from the slight rotations or wobbles of StratoSats in windy conditions. Especially in the scenario using ultra-massive (UM)-MIMO  antenna to communicate in THz band, the channels present the unique triple delay-beam-Doppler squint effects, which needs specialized channel estimation technologies for effective compensation. The technique will be further introduced in Section 3.
	The reliability of beamforming deteriorates due to two main reasons. Firstly, the long distance of the link combined with the extremely narrow beamwidth results in significant degradation of communication quality due to even minor deviations in angle parameters during positioning. Secondly, both balloons and airships operate in quasi-static conditions with small propulsion systems, making their positions and orientations subject to random disturbances from the environment. Even slight rotations of the antenna array mounted on the transceiver can lead to sub-optimal beam pointing. Such complex scenarios are not encountered in airborne or spaceborne networks, rendering existing solutions ineffective. To cope with this issue, SoftBank Corporation have devised a footprint fixation technology. This technology involves adjusting the cylindrical phased-array antennas of the balloon to mitigate the effects of potential motion and rotation of StratoSats, thereby ensuring the stability of Internet coverage.
	
	\color{black}
	\section{Key Technologies of NS-COM network}
	Having explored the disparity among different layers and NS-COM network's noteworthy advantages, our focus turns to underscore the significance of investigating the key technology of NS-COM network. This segment provides a detailed examination of its pioneering efforts in channel modeling, random access, channel estimation, array-based beam management and joint network optimization. Understanding these key technological aspects is crucial for unlocking the full potential of StratoSat and advancing the capabilities of its communication system.
	
	\subsection{Channel Modeling}
	Channel modeling is a crucial aspect of communication system design as it involves creating mathematical representations of the transmission medium to predict signal behavior. In the context of NS-COM network, the challenge lies in the unique and dynamic nature of the near-space channel. Unlike traditional terrestrial channels, the near-space environment introduces factors such as variable atmospheric conditions, high altitudes and platform mobility. These factors make accurate channel modeling in NS-COM network particularly challenging. The need to account for the intricate interplay of atmospheric effects, platform movements and other environmental variables complicates the development of precise models.
	
	To further boost signal transmission distance and quality, the idea of introducing multiple-input multiple-output (MIMO) has been put on the agenda \cite{9961131}. MIMO technology deploys multiple antenna elements at both the transmitter and receiver, significantly enhancing the performance of wireless communication systems in complex multipath environments without increasing transmission power or bandwidth \cite{9849060}. Enabling MIMO technology in NS-COM network is a promising solution for improving spectral efficiency. 
	\color{black}{Therefore, investigating the three-dimensional (3D) StratoSat-MIMO channel model has become an indispensable aspect in NS-COM communication research. In \cite{6136823}, a cooperative MIMO channel model was introduced and the spatial correlation in multiple scenarios was examined. However, these models were designed for two-dimensional (2D) channel models, neglecting the elevation angle. To address multi-user scenarios more comprehensively, the authors of \cite{5409569} introduced a 3D StratoSat-MIMO channel model. It considered the non-stationary properties of multi-user MIMO channels, accounting for the appearance and disappearance of scatterers. The significance of including moving scatterers in propagation models was demonstrated in \cite{6708467}. In \cite{821698}, a birth and death process models the appearance and disappearance of scatterers but fell short in modeling their reappearance behavior. The authors of \cite{7779116} utilized an $M$-step Markov process to portray the dynamic evolution of scatterers, however it only considered stationary scatterers. In \cite{8171024}, both the mobility and the evolution of scatterers were considered by incorporating an M-step 2-state Markov process into the proposed channel model.
	}
	\color{black}
	\subsection{Random Access}
	
	The mMTC refers to the ability of a communication system to handle a large number of user devices simultaneously accessing the network \cite{c2,9726781}. This concept is particularly relevant in the context of the Internet of Things (IoT) and  the upcoming 6G communication networks, where a vast number of devices, such as sensors, smart devices and machines, require connectivity \cite{c1}. The application of random access techniques is crucial to efficiently manage the simultaneous access of multiple users competing for limited resources, especially in the NS-COM network assisted SAGSIN \cite{10123361}. The necessity arises from the challenges posed by a massive number of users attempting to access the network concurrently \cite{9487496,9726781}.

	Grant-free sourced random access involves devices transmitting non-orthogonal pilot sequences alongside payload data to the BS for active device detection (ADD) and channel estimation before coherent data detection \cite{m1}. It involves the transmission of data without authorization \cite{8961111}. Sparse device activity in URLLC inspired compressed sensing-based joint ADD and channel estimation design, with solutions addressing single-antenna limitations but requiring adaptation for multi-antenna systems \cite{8612905,9266124,6196146}.
	The authors of \cite{8612905} introduced a coherent detection framework utilizing non-orthogonal pilots, aiding in ADD and channel estimation.
	In \cite{9266124}, the authors emphasized the sporadic nature of uplink traffic, where only a limited number of devices activate simultaneously. By proposing dimension reduction techniques for joint ADD and channel estimation, the authors of \cite{6196146} focused on reducing computational complexity.
	
	Grant-free unsourced random access relies on a common codebook-based non-coherent detection framework. In contrast to sourced random access, in this scenario, the BS is focused solely on estimating a series of transmitted messages without concern for the identities of the transmitters. 
	In \cite{8006984}, a common codebook-based non-coherent detection framework was introduced. It emphasizes the estimation of transmitted messages without identifying specific transmitters.
	The authors of \cite{8006985} introduced a low-complexity coding scheme that divides the transmission period into sub-blocks, allowing each device to randomly choose a sub-block for transmission, thereby reducing complexity.
	These approaches struggle with issues related to codebook size, computational complexity and payload efficiency in addressing stringent latency and reliability requirements of massive URLLC. Integrating these random access paradigms within a unified detection framework remains a challenge, necessitating a shared random access procedure and transceiver hardware design to optimize performance without additional hardware complexity or cost. The authors of \cite{ke} designed a unified semi-blind detection framework for grant-free sourced and unsourced random access to better enable next-generation URLLC. In \cite{add3}, a StratoSat identified as a grant-based user sought to access a satellite network concurrently with multiple earth stations termed as grant-free users, using NOMA-assisted semi-grant-free protocol. Two uplink transmission schemes for both perfect channel state information (CSI) and imperfect CSI cases were proposed to achieve higher sum rate.


	\subsection{Channel Estimation}
	Channel estimation is a crucial process in communication networks that involves predicting the characteristics of the transmission medium \cite{9452036}. The unique dynamics of the near-space environment, including variable atmospheric conditions and platform mobility, necessitate accurate channel estimation for the subsequent signal transmission.
	
	Ensuring the QoS necessitates the acquisition of reliable CSI at the transceiver \cite{8624276}. However, although StratoSat operates in a position close to stationary relative to the Earth, which makes the CSI less capricious than satellites, this task is still particularly challenging due to the inherent wobbling of aerial BSs. The high-speed mobility of aerial signal receivers such as flying aircraft/UAVs also adds to the difficulty when experiencing rapid changes in their positions and orientations. Consequently, aerial communication links exhibit rapidly time-varying fading characteristics, posing considerable difficulties in achieving accurate channel estimation and tracking.

	Several channel estimation and tracking schemes in \cite{9398858,9069646} have been proposed to obtain accurate estimates of fast time-varying channels while reducing the training overhead caused by frequent channel estimation. The authors of \cite{9398858} introduced an efficient channel estimation and tracking scheme aimed at resolving the performance degradation issue arising from the distinctive triple delay-beam-Doppler squint effects observed in aeronautical THz channels. A blind channel estimation and equalization method to address the distortion induced by multipath and fading in an aeronautical telemetry channel was proposed in \cite{9069646}. Utilizing the property that the variations of angles of arrival/departure (AoAs/AoDs) are much slower than that of path gains, the authors of \cite{9069646} used staged estimation on the aforementioned dominant channel parameters to reduce time overhead and provide fast channel tracking.

	\color{black}
	\subsection{Array-Based Beam Management}
	
	
	In the NS-COM network, array-based beam management plays a vital role, particularly with the use of high frequencies like millimeter-wave for ground-to-air and THz for air-to-air links. These high frequencies offer increased data rates but face signal attenuation issues. To tackle this, antennas generate beams with heightened gain. Precise control of these beams is crucial for effective aerial communication.
	StratoSat's susceptibility to orientation variations due to motion and airflow in the troposphere and stratosphere is another factor. Beam management must ensure continuous alignment to counter these changes. Regular monitoring and real-time adjustments of the narrow beams are necessary to maintain precise alignment, compensating for platform movements \cite{10179219}. This capability ensures reliable communication links, vital for NS-COM network stability in near-space environments.
	
	While StratoSats tend to remain relatively stationary concerning the Earth, users may still move, especially in scenarios involving communication between airships and satellites. However, during aerial propagation, there is an elongated delay, leading to channel aging issues \cite{5067722}. As a result, beam tracking becomes essential to address these challenges. By employing sophisticated array antenna systems, these platforms can effectively steer and shape these narrow beams to precisely target and serve the designated areas, optimizing the communication link quality and ensuring reliable data transmission \cite{7928493}. 
	
	When considering antenna arrays on StratoSats, several factors come into play. One of the primary challenges is the limited size and weight constraints of the StatoSats, which make it difficult to implement large and sophisticated antenna arrays. As a result, fully-digital arrays with a large number of elements may not be feasible due to the added weight, power consumption and hardware cost. These factors could significantly impact the flight duration and overall mission capabilities \cite{9598918}.
	On the contrary, hybrid analog-digital arrays have been proposed and deployed on aerial vehicles \cite{9099976,10024899}. These arrays combine analog beamforming techniques with digital signal processing, enabling them to achieve beamforming capabilities while minimizing the number of RF chains required \cite{10195164}. This approach not only reduces power consumption and hardware cost but also mitigates the difficulties associated with fully-digital arrays.

	\subsection{Joint Network Optimization}
	
	The introduction of introducing NS-COM network into SAGSIN presents challenges in fostering collaboration among heterogeneous networks and within the NS-COM network itself, particularly concerning joint beamforming and edge computing.
	
	Collaborative beamforming across diverse network layers enhances optimization flexibility, thereby elevating network throughput. Rolling out additional StratoSats within specified hotspot zones involves multiple network nodes utilizing identical frequency bands simultaneously. To enhance network data rates, a proposed solution in \cite{9869727} introduced an iterative approach integrating precoding and band allocation algorithms tailored for multi-antenna StratoSats. These methods aim to optimize data rates and employ various band allocation and beamforming techniques.
	The objective of the study presented in \cite{10304301} centered on optimizing network-wide throughput. This optimization considers constraints related to StratoSat payload connectivity, power limitations of StratoSat and BSs, and backhaul limitations. The study aimed to jointly determine user-association strategies and associated beamforming vectors for each user. To accomplish this, an iterative modular approach was employed to solve a mixed-integer optimization problem.
	The authors in \cite{8644171} emphasized using energy-harvesting relays aiding data transmission from a multi-antenna HAP to a distant receiver within a cellular system. It aimed to maximize throughput under  interference constraints at cellular user equipments. The investigation carried out by the authors in \cite{8255106} centered on minimize interference originating from adjacent StratoSats in a multiple user equipment (UE) scenario. Each UE's beamforming vector was determined by maximizing its signal to pilot contamination ratio, then by using solely local statistical channel details, the optimization of signal to pilot contamination ratio is completely independent across UEs.
	
	Additionally, a globally supported edge computing approach can effectively assist in joint decision-making for beam deployment, resource allocation and interference mitigation \cite{m1}.
	Micro edge servers can be introduced owing to some recent research that leveraged UAVs as aerial platforms carrying edge servers to provide various services and support ground BSs or directly connect devices. For instance, in [80, 81], the authors established an air-ground integrated edge computing network where UAVs can be flexibly deployed and scheduled to enhance communication, caching and computing in wide-area networks.
	The StratoSat enabled overloaded terrestrial BSs to offload some computing tasks to the edge servers carried by StratoSats. This offloading alleviates the load on terrestrial
	BSs, allowing devices that cannot access them to establish
	direct connectivity with NS-COM network \cite{9615110}.

	\section{Typical Application Scenarios}

	Due to StratoSat's quasi-stationary positions relative to the Earth and its proximity compared to satellites, these airships offer unique opportunities and hold great promise for revolutionizing both communication, sensing and many other domains, making them a promising platform for various endeavors. The forthcoming sections will delve deeper into the specific applications and technical considerations related to the utilization of StratoSat in various communication and sensing scenarios, further elucidating their potential impact on modern space technology.
	
	\subsection{Structural Expansion in SAGSIN Communication}
	
\begin{figure*}[t]
		\centering
		\includegraphics[width=1\textwidth]{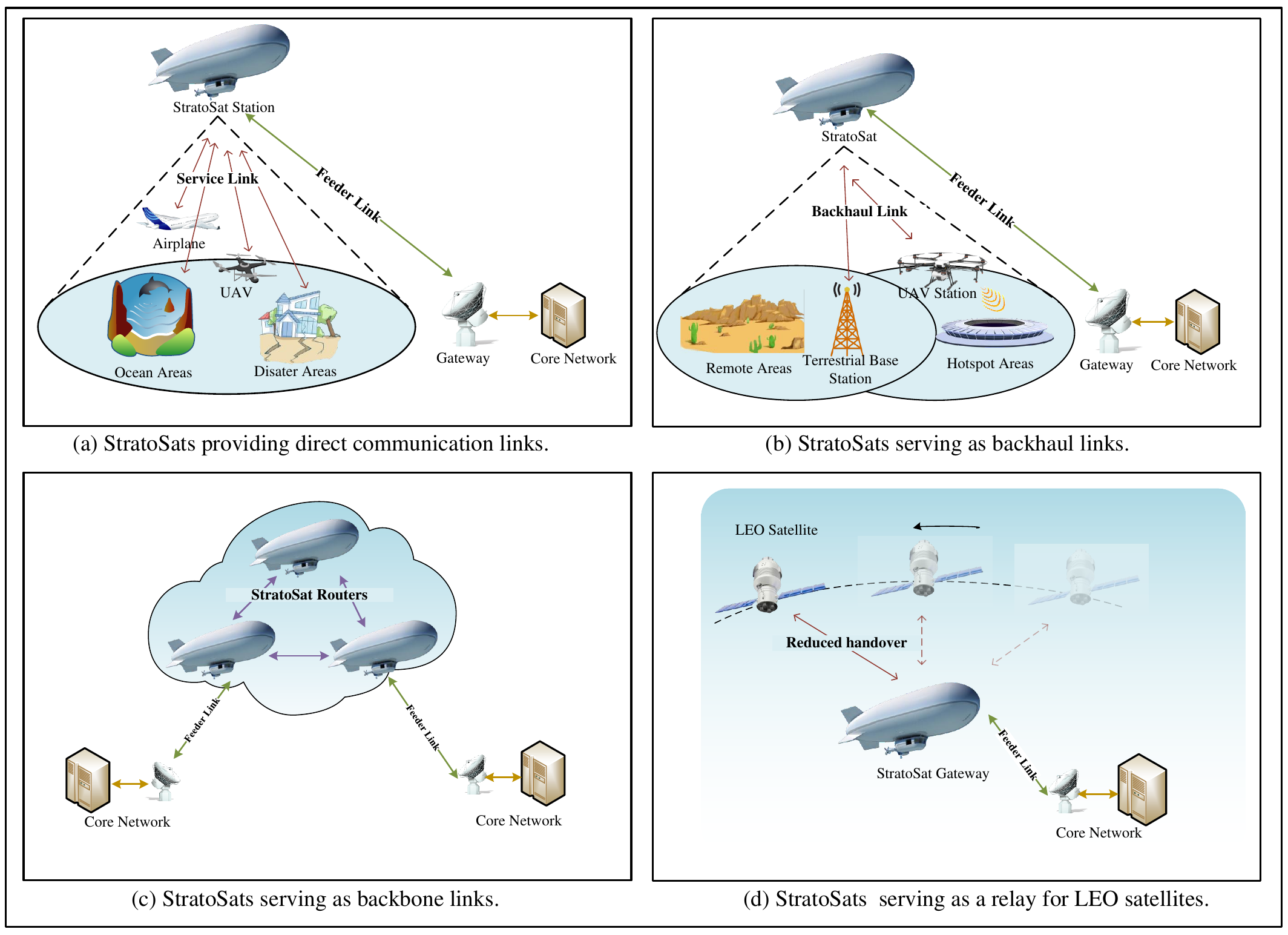}
		\caption{A schematic diagram of different roles StratoSats play in the network architectures.}
		\label{fig:2}
	\end{figure*}

	StratoSats hold tremendous promise in shaping the future network architecture due to their expansive coverage range, stable channel conditions and quasi-stationary properties, as such can play important roles in wireless access, backhaul link, backbone transmission, stable interface for satellites, which are exhibited in Fig. \ref{fig:2}.
	
	StratoSats play a crucial role in providing diverse communication services to users in disaster and underserved areas (Fig. \ref{fig:2} (a)). Stratospheric access points extend the reach of existing terrestrial network, bridging coverage gaps and enhancing connectivity. Aerial and sea users, including aircraft, ships and UAVs, benefit from seamless communication while moving across different regions or over open waters, thanks to StratoSats acting as floating BSs. 
	A notable example of this capability is Project Loon by Google's parent company Alphabet, where stratospheric balloons equipped with communication payloads provided internet access to remote and disaster-affected regions \cite{8256905}. The project showcased the potential of StratoSats in narrowing the digital divide and connecting underserved communities.

	StratoSats can also serve as valuable backhaul links in the network architecture (Fig. \ref{fig:2} (b)). Positioned as communication nodes above the Earth's surface, they enable direct and efficient communication between terrestrial BSs and the core network. In remote regions, where laying fiber-optic cables or other wired infrastructure can be challenging or expensive, StratoSats offer a feasible and flexible solution for extending backhaul connectivity.
	In this capacity, StratoSats complement existing terrestrial backhaul infrastructure, particularly in areas with geographical barriers, sparse population, or difficult terrain. Also, for hotspot areas, StratoSats can efficiently handle a large number of user connections simultaneously, easing congestion in densely populated areas and crowded events. A noteworthy real-world example of this application is Project Taara, a subsidiary of Alphabet, which utilized laser technology to establish high-speed, wireless optical communication links between terrestrial BSs in remote locations \cite{taara}. This project successfully demonstrated the feasibility of using StratoSats to create backhaul connections, bypassing the need for physical cables and reducing the cost and complexity of network deployment in challenging environments.
	
	StratoSats offer a unique advantage as backbone links in network architecture (Fig. \ref{fig:2} (c)). They can act as reliable relay nodes, establishing connections among core network hubs and data centers, ensuring seamless data transfer and communication services even in challenging geographical locations or areas with limited terrestrial network coverage.
	One significant advantage of utilizing StratoSats as backbone links is their potential to support advanced transmission technologies like mmWave and THz communications, as well as FSO \cite{8255764, 9374451}. These high-frequency bands and optical communication methods can enable extremely high data rates, allowing for ultra-fast and high-capacity data transfer over long distances, thereby enhancing the overall network performance.
	
	Another crucial role that StratoSats can play in future networks is serving as a relay for LEO satellites (Fig. \ref{fig:2} (d)). As LEO satellites move at high speeds in their orbits, they can cause frequent disconnections and handover issues when communicating with ground-based terrestrial gateways \cite{10110015,9928043}. StratoSats' high-altitude placement allows them to have a broader field of view, enabling a single StratoSat to simultaneously handle communications with multiple LEO satellites within its footprint. As a result, the handover problem can be effectively managed. Moreover, the stratospheric environment exhibits relatively stable atmospheric conditions, resulting in reduced signal attenuation and interference compared to ground-based systems, which further enhances the reliability and efficiency of communication with LEO satellites.

	\color{black}
	\subsection{Civil Aviation Communication}
	
	Currently, passengers onboard aircraft lack internet connectivity, and even if available, it requires satellite access, which can be facilitated via a StratoSat connecting to the terrestrial core network. While terrestrial flight routes can utilize ground-based stations, maritime routes rely solely on satellite communication. StratoSats can extend communication services to aircraft over maritime routes and polar routes.
	In addition to catering to passenger needs, StratoSats can also benefit other stakeholders in the aviation industry, such as airlines, aircraft manufacturers, and air traffic control authorities. By facilitating seamless communication between aircraft and ground-based operations, StratoSats can improve flight coordination, enhance safety measures, and optimize airspace utilization. 
	However, these scheduling signals require frequent communication and have a wide communication range. Therefore, new technologies need to be applied to StratoSat to meet the higher quality and spectral efficiency communication requirements.
	
	THz communications enable the deployment of Ultra-Massive MIMO (UM-MIMO) transceivers with tens of thousands of antennas. This facilitates effective mitigation of the substantial path loss experienced by THz signals, thereby extending the communication range even further.
	The THz frequency band, ranging from 0.1 THz to 10 THz, is anticipated to offer substantially wider bandwidths compared to mmWave, enabling support for ultra-broadband in the range of tens of  GHz and ultra-high peak data rates in the Terabit per second range. The intrinsic suitability for near-space environments characterized by low path loss makes it a promising candidate for 5G cellular networks, wireless local area networks, and wireless personal area networks \cite{9398858}, thereby underscoring the significance of exploring THz communication channels.
	
	\begin{figure*}[t]
		\centering
		\includegraphics[width=1\textwidth]{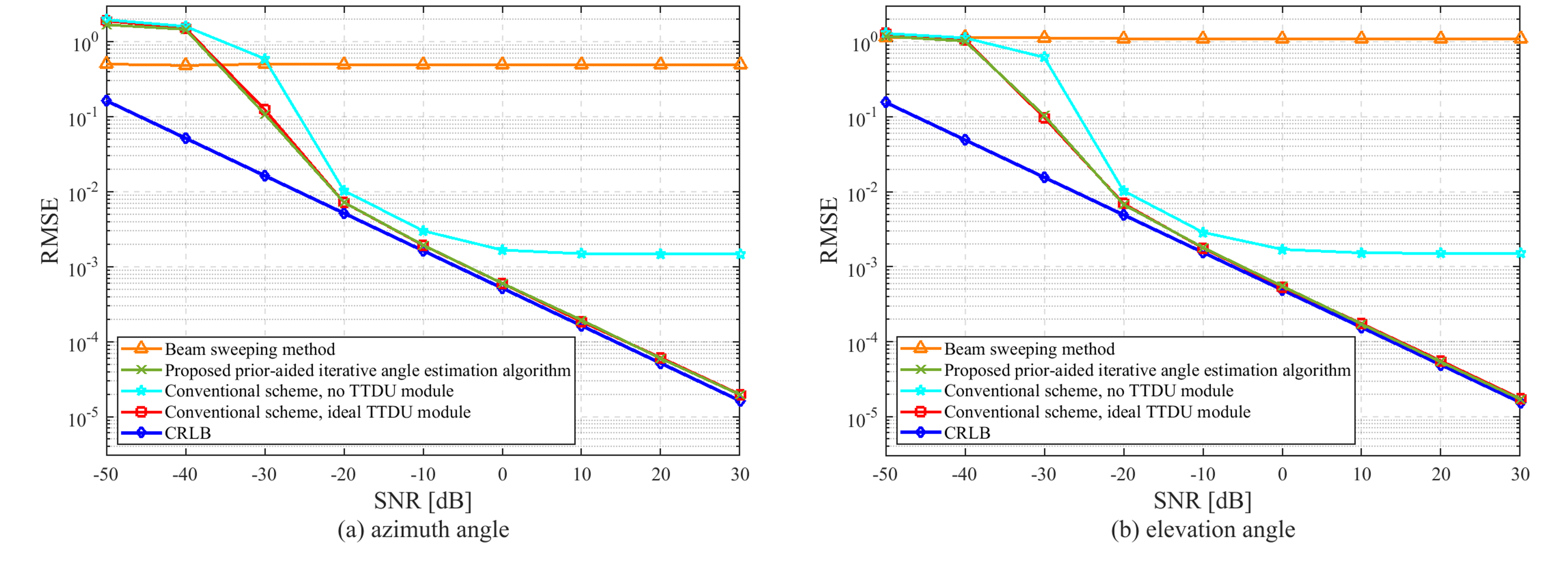}
		\caption{\color{black}RMSE comparison of angle estimation: (a) azimuth angle; and (b) elevation angle in \cite{9398858}.}
		\label{fig:10}
	\end{figure*}
	
	Although THz communication brings great advantages, it also poses high demands on technology \cite{9941044}. When the direction of arrival deviates from perpendicular to the array, varying propagation delays are observed across different antennas receiving the same signal within the array aperture. This delay discrepancy can extend to multiple symbol periods, especially in the context of ultra-broadband THz communications. Traditional analog beamforming employs phase shifters designed according to the center frequency of the bandwidth, resulting in severe beam squint effects in THz UM-MIMO systems with ultra-broad bandwidth. To compensate the effect, the authors of \cite{9398858} introduced an angle estimation for a complex THz band StratoSat-to-aircraft link. Through the integration of a true-time delay unit (TTDU) module prior to the analog beamformer, a frequency-dependent phase-shifter network was developed to achieve optimal beam alignment across the complete communication spectrum. This configuration effectively alleviated the delay squint effect and realized a frequency-dependent phase-shifter network to ensure desirable beam.
	Fig. \ref{fig:10} compares the root mean square error (RMSE) performance of the aforementioned angle estimation with the following benchmarks. The labels ``beam sweeping method'' is the algorithm presented in \cite{56}. The labels ``no TTDU module'' and ``ideal TTDU module'' indicate the transceiver adopting ideal TTDU module and without considering TTDU module, respectively. The label ``conventional scheme'' indicates directly applying the conventional two-dimensional unitary estimation of signal parameters via rotational invariance techniques (TDU-ESPRIT) algorithm to estimate angles as those used in existing mmWave systems \cite{8846224}. From the simulation result, it can be seen that the RMSE curves of ``proposed algorithm'' and ``conventional scheme'' using ``ideal TTDU module'' almost overlap, and they are very close to the Cramer-Rao lower bounds (CRLBs) of azimuth and elevation angles at high signal-to-noise ratio (SNR).

	\color{black}
	
	\subsection{Remote and Urgent Communication}
	
	StratoSat has a vast range of applications in the field of communication, including service for users with obstructed ground signals, emergency service and high-speed rail communication.
	
	\textbullet \textbf{Service for Users with Obstructed Ground Signals}:
	While terrestrial data transmission is more fragile to shadowing \cite{Xiao}, StratoSats are essential to provide extra reliable LoS links, while the terrestrial data transmission is more fragile to shadowing, StratoSats are essential to serve ground BS edge users due to interference, handover problems, and low signal power at the cellular edge, rural area or other area that lacks LoS link to the BS \cite{9356529,8746491}. The challenges of deploying ground BSs in such areas, along with terrain limitations, make StratoSats a cost-effective and versatile solution, offering improved coverage and communication quality without infrastructure constraints.
	
	\textbullet \textbf{Emergency Service}:
	The deployment of StratoSat in disaster-prone regions, such as those affected by tsunamis, hurricanes and earthquakes, can be achieved with relative ease and cost-effectiveness. This deployment empowers these areas with reliable and accessible Internet connectivity. Disaster-stricken regions require prompt completion of rescue operations and post-disaster reconstruction \cite{9526159}. Therefore, there is a high demand for timely deployment of communication solutions. The exceptional maneuverability of StratoSat enables them to traverse vast distances while seamlessly transitioning between dynamic movement and strategic hovering over specific locations. This unique capability makes StratoSats highly versatile and effective in fulfilling various communication and observation tasks.

	\textbullet \textbf{High-Speed Rail Communication}:
	In recent years, as railways have undergone rapid growth, there has been a growing desire among passengers to have seamless wireless Internet access through wireless local area network technology \cite{train}. However, if the conventional cellular mobile system is employed, the signal would need to frequently switch between different cellular networks as the train moves along the railway \cite{7897321,1593555}. Since StratoSats serve a much larger area than the terrestrial network, frequent cell switching can be avoided.

	\subsection{Weather Monitoring}
	
	Near-space has unique physical characteristics, such as high radiation, extremely low temperature and low pressure. In recent years, the role of near-space environment detection has become increasingly prominent, especially in the field of climate change monitoring \cite{9996809}. Conducting in-situ real-time observations of near-space is essential for studying, monitoring and forecasting the Earth's atmosphere, as well as for studying and predicting meteorological disasters \cite{7190585}. StratoSat's maneuverable control and strong payload capacity make it well-suited for carrying equipments such as radars and radio sondes. The distributed deployment of radio sondes can bring higher resolution and precision to the measurement results.

	For example, typhoon is one of the most severe natural disasters in the world, occurring over 80 times annually and often causing devastating destruction to coastal cities \cite{jiang2018deep}. Additionally, establishing a typhoon model can help people understand the wind and current fields over the ocean, aiding in navigation planning for ships. However, the internal circulation structure of a typhoon is complex, with strong winds and turbulence, researchers have limited knowledge of the structural evolution mechanism of the typhoon's core region. Moreover, typhoons are mainly caused by the exchange of energy between water vapor on the sea surface, making it difficult for shore-based detection stations to obtain data on sea conditions \cite{9996809}.
	\begin{figure*}[h]
		\centering
		\includegraphics[width=0.8\textwidth]{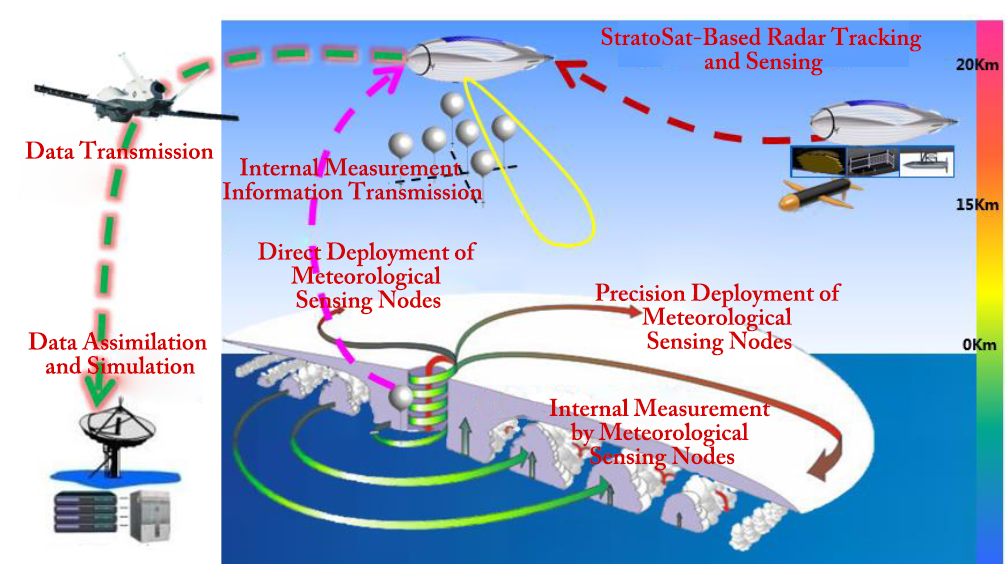}
		\caption{The process of StratoSat tracking and detecting typhoons.}
		\label{fig:typhoon}
	\end{figure*}
	For the above reasons, typhoon detection carried out by high-pressure balloons carrying floating meteorological sensing nodes in conjunction with StratoSats has become a valuable resource for studying typhoon structure evolution and predicting typhoon trajectories. 
	The workflow of an actual measurement is meticulously designed as follows and also illustrated in Fig. \ref{fig:typhoon}. SrtaotSats, guided by typhoon tracking flight controllers, persistently follow the typhoon's path. Utilizing onboard radars for beam scanning, they closely monitoring the typhoon's three-dimensional wind field and structural nuances. Then, strategic deployment of floating meteorological sensing nodes, either directly or via precision deployment devices, allows for detailed data collection near the typhoon eye. The collected data, comprising temperature, humidity, wind, and pressure readings, is transmitted back to the airships and then relayed to a ground data center, where a four-dimensional, high-resolution digital typhoon simulation system integrates and simulates the data, enhancing the accuracy of typhoon forecasting and contributing to a deeper understanding of these complex natural phenomena \cite{tf}.
	This integrated approach allows for more comprehensive and accurate data collection, contributing to a better understanding of typhoons and improving forecasting capabilities.
	\subsection{Carbon Neutrality}
	\begin{figure*}[h]
		\centering
		\includegraphics[width=0.8\textwidth]{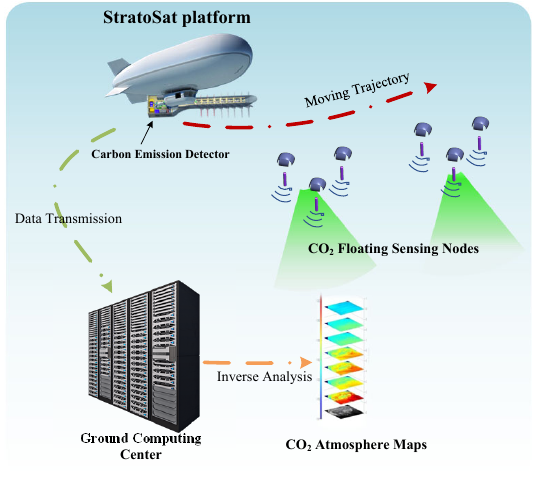}
		\caption{A schematic diagram of StratoSat based atmosphere $\text{CO}_{2}$ detection system.}
		\label{fig:4}
	\end{figure*}
	Near-space environment detection can also address the critical issue of gas content monitoring. Greenhouse gases, of which carbon emissions are a key component, are a major contributor to global warming. Therefore, it is necessary to accurately monitor and analyze carbon emissions to effectively control the rate of global temperature rise \cite{wang2021technologies}. In-situ monitoring is considered the most reliable and accurate method for this purpose, as it enables direct measurements of emissions at the source \cite{wu2022review}.
	
	One promising approach to achieve high spatiotemporal resolution carbon emission maps is through the deployment of floating sensing node balloons at lower altitude compared to StraotSats, as shown in Fig. \ref{fig:4}. Equipped with advanced sensors that can detect various types of greenhouse gases, including carbon dioxide, methane and nitrous oxide, these balloons can capture emissions from various sources such as industrial facilities, transportation and natural phenomena like wildfires. The balloons use atmospheric circulation and wind field prediction techniques for flight trajectory maintenance and control, maintaining the ability to provide 3D spatially layered measurements, enabling a more comprehensive understanding of carbon emissions. By combining data from multiple balloons at different altitudes and performing data inversion, researchers can create a detailed map of the distribution and concentration of greenhouse gases in the atmosphere, which can help identify and regulate the sources of emissions. This technology not only facilitates monitoring of carbon exhaustion from industrial facilities, but also enables high-precision detection of forest fires, thereby contributing to forest fire prevention.
	
	Compared to previous 2D carbon emission maps, the use of balloon detection nodes has the advantage of detecting emissions and leaks in key areas such as petroleum and factories. The balloons can fly over these areas and collect data, which can help authorities and industry regulators monitor compliance and identify potential violations. In conclusion, near-space environment detection has the potential to revolutionize our understanding and management of greenhouse gas emissions. The deployment of floating sensing node balloons and balloon detection nodes can provide a more accurate and comprehensive picture of carbon emissions, which can inform policy decisions and guide actions to mitigate climate change.
	
	\section{Open Issues and Future Prospects}
\begin{figure*}[h]
		\centering
		\includegraphics[width=0.8\textwidth]{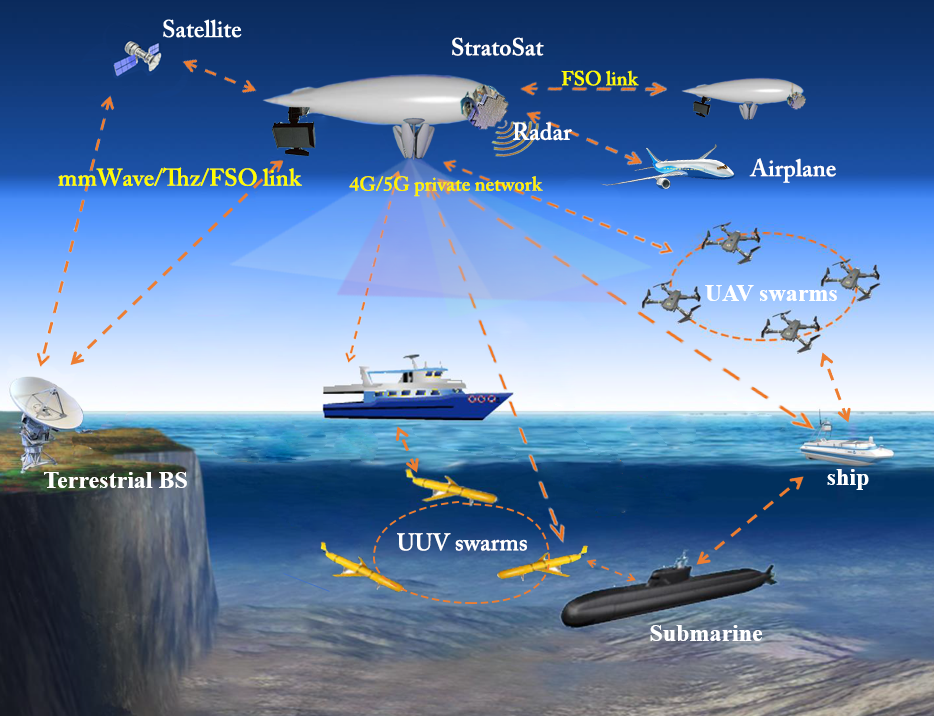}
		\caption{\color{black}Schematic diagram of StratoSat's potential application scenarios with future technologies.}
		\label{fig:5}
	\end{figure*}

	Despite its potential, the research of applying other communication technologies in NS-COM is still in its infancy, many key research issues are still open. In this section, we investigate several candidate communication techniques for NS-COM, followed by a discussion about the potential research opportunities therein. The application scenarios of some of these technologies are illustrated in Fig. \ref{fig:5}.
	\subsection{StratoSat-to-Ground Direct Links for Mobile Terminals}
	While satellite communications often encounter cell switching issues and beam alignment \cite{directconnect,xu2023enhancement,arr}, StratoSats remain relatively stationary with respect to the ground, eliminating the need to consider cell switching or rapid channel variations, also resulting in minimal time and frequency offsets between the transceiver parties.
	A StratoSat-to-ground direct link is demonstrated in Fig. \ref{fig:5}. It uses mobile devices such as smartphones, which presents challenges in terms of hardware limitations, antenna gain, access protocols and data frame structures. The hardware limitations of mobile phone antennas result in a lower frequency band for direct communication with StratoSats, which can easily cause interference with existing ground-based mobile communication networks. Additionally, limited antenna gain and a restricted number of RF chains in mobile devices constrain communication performance, indicating that the hardware capabilities need further improvement. Researchers are exploring innovative antenna designs and hardware improvements to enhance mobile phone capabilities for NS-COM. They are also developing efficient access protocols and handover procedures to optimize communication between mobile devices and satellites. Despite the complexities, the potential benefits of reliable connectivity in remote areas drive ongoing research and development in this field. As advancements continue, StratoSat-to-ground direct link holds promise for revolutionizing global communication systems.
	
	\subsection{Emerging Antenna Design: Reconfigurable MIMO and Holographic MIMO}
	
	Reconfigurable MIMO acts as transmission terminal to increase the degree of freedom of beam pattern \cite{ykk,reconfigurable1,reconfigurable2, reconfigurable3}.
	A novel concept of utilizing reconfigurable MIMO has been introduced to unlock the untapped potential of the electromagnetic realm, aiming to amplify the efficiency of information transmission \cite{ykk}. 
	Reconfigurable MIMO can be introduced as terrestrial BS antennas to communicate with StratoSats \cite{wu2023deep}. Additionally, reconfigurable MIMO has demonstrated  its ability to actively manage the transmission environment \cite{RIS1,RIS2}.
	Since this technique has just been introduced in NS-COM, the pattern reconfiguration that aims to actively manage the transmission environment for NS-COM still needs to be modified and trained. The optimization objective can aim at redistribute power in signal directions, enhancing interference suppression and reinforcing energy efficiency.
	\color{black}
	Estimating the full CSI at once is not feasible for reconfigurable MIMO due to the long communication distance of StratoSat, which results in longer signal transmission times. The prolonged channel estimation training process may lead to outdated CSI, potentially affecting the accuracy of data detection and the efficiency of precoding. On the other hand, exploiting reconfigurable capabilities could introduce randomness into the channel, which might be advantageous for implementing covert communication strategies on StratoSat, such as channel randomization or radiation pattern hopping.
	
	A holographic MIMO involves incorporating a vast quantity of small and cost-effective antennas or reconfigurable elements into a confined area, achieving a holographic array with a continuously spatial aperture \cite{9374451}.
	However, the physical channels associated with holographic
	MIMO cannot be represented in terms of the finite-dimensional matrices. The further channel estimation and beamforming technology, requires abundant double integral calculation due to the continuity of the antenna surface, leading to extremely high computing performance requirements. Thus, it is imperative to investigate methodologies for integrating high-performance computing devices within the constrained payload capacity of StratoSats.
	\color{black}
	
	\subsection{Federated Learning in NS-COM Networks}

	In the SAGSIN, utilizing federated learning in StratoSat as distributed edge servers offers significant merits. Unlike traditional centralized approaches, federated learning allows StratoSat to directly learn a model for specific tasks, such as trajectory control, resource scheduling, visual recognition and beamforming, without the need to collect data from all participating nodes. StratoSat can collaboratively update the model parameters without sharing raw data, ensuring that sensitive information is not compromised.
	\color{black}
	Given that StratoSats might serve various regions or user communities, there's a critical need for efficient resource distribution and scheduling techniques to oversee federated learning operations. This includes optimizing the allocation of computational, communicative, and storage capacities to satisfy diverse regional or user requirements.
	But since the energy supply for StratoSat may be limited, it might be difficult to meet the high-energy demands of federated learning. On the other hand, due to StratSat's mobility, nodes may be temporarily added or removed, requiring adaptability and scalability to cope with changing environments and demands. Flexible and scalable federated learning algorithms and frameworks also need to be designed to meet the needs of different scenarios and applications.
	Moreover, federated learning requires the exchange of a large amount of model parameters and gradient updates between participants, leading to significant communication overhead. Therefore, it is necessary to develop a gradient transmission policy with higher energy efficiency and lower latency.
	\color{black}

	
	\subsection{Maritime Communication}
	StratoSat's versatility extends to providing communication services in maritime environments. Acting as a communication relay, StratoSats establish seamless links between ships, submarines, onshore facilities and aircrafts, especially in areas lacking terrestrial BS service,  exhibited in Fig. \ref{fig:5}. This enables reliable and efficient communication for maritime operations and safety, addressing challenges posed by vast distances and natural barriers in sea regions. StratoSat's deployment offers a transformative solution, ensuring continuous and effective connectivity for ships and onshore facilities, enhancing maritime communication and coordination.
	
	However, the maritime communication through NS-COM also encounters several challenges. Firstly, the overlapping coverage of different nodes and ships in the ocean results in strong spatial correlations, leading to significant inter-user interference. Secondly, the multitude of nodes in these areas generates a high demand for multi-modal data transmission, while the available bandwidth resources may not be sufficient to meet these requirements. Thirdly, the practical constraints of StratoSats on payload, size and hardware power consumption  limit the downlink transmission rates for maritime users. Moreover, hardware limitations in maritime terminal sensors also restrict the capacity of air-sea communication links.
	
	\color{black}
	
	\subsection{Electromagnetic Spectrum Sensing and Adversarial Game}
	Electromagnetic spectrum sensing and adversarial game ensure the efficient utilization of radio spectrum resources, protect communication networks from malicious interference and attacks, and support military operations by jamming enemy communications while securing one's own networks. Additionally, they play a crucial role in monitoring and identifying radio spectrum activities for intelligence gathering and surveillance. Fig. \ref{fig:5} is a simplified illustration of electromagnetic spectrum sensing and adversarial game.
	In spite of the advantages, these applications face technical challenges. The dynamic changes in the spectrum environment requires real-time monitoring and adjustments. The applications also require  efficient algorithms and data processing techniques to manage large spectrum data volumes. Additionally, there is a necessity to accurately assess the effectiveness of countermeasures. Another challenge is ensuring electromagnetic compatibility to prevent interference with legitimate communications. There is also challenge in addressing energy constraints on near-space platforms to design energy-efficient systems for extended operational endurance.
	
	\subsection{Integrated Sensing and Communications}
	The integrated sensing and communications system onboard the StratoSat enables real-time data transmission and processing, facilitating rapid acquisition and transmission of various detected environmental data and perception information. This capability provides timely monitoring and response capabilities for ground users or other systems. The integrated sensing and communications system can consolidate multiple sensors, enhancing StratoSat's perception capabilities of the surrounding environment. Furthermore, guided by radar data, the system can improve the physical layer transmission algorithms to enhance communication quality \cite{9898900}. 
	Despite these advantages, the limited space onboard the StratoSat may lead to densely packed communication equipment and radar sensing devices, increasing the risk of electromagnetic interference. Such interference could potentially compromise the performance and imaging quality. Therefore, measures must be implemented to mitigate or suppress electromagnetic interference. Moreover, integrated sensing and communications system requires addressing compatibility and coordination issues between different technologies to ensure overall performance and efficiency. The complex environment surrounding the StratoSat makes it challenging to anticipate reasonable expectations for radar perception accuracy and range. Thus, exploring how to balance perception indicators of the radar and communication efficiency is imperative.
	
	\subsection{StratoSat-Based Radar Detection and Imaging}
	The StratoSat-based radar detection and imaging system is demonstrated in Fig. \ref{fig:5}, which enables remote monitoring and reconnaissance over a wide coverage area, suitable for monitoring vast ground regions or ocean surfaces, even under complex terrain conditions. Unlike optical imaging, radar detection and imaging is not affected by weather conditions such as cloud cover or rain, thus providing stable and reliable monitoring services even in adverse weather conditions.
	However, due to the continuous operation required for radar to detect unexpected objects, the limited energy supply on StratoSat might be unable to sustain long-term radar sensing and imaging tasks. Therefore, energy-efficient radar systems need to be designed, and the energy management of the StratoSat must be optimized to extend mission duration. Additionally, the suppression of interference and environmental noise is critical to maintaining the integrity of imaging quality and accuracy, affecting the system's ability to identify and locate targets. The ionospheric radiation above the StratoSat may also distort radar signals.
	
	\subsection{NS-COM Assisted Enhanced Global Navigation System}
	NS-COM assisted enhanced global navigation system can provide high-precision positioning, including indoor and complex environment positioning, along with extending the coverage range of navigation system and ensuring reliable positioning services even in signal-obstructed areas like densely populated urban zones or mountainous regions. Furthermore, NS-COM facilitates real-time data transmission, enhancing the navigation experience by enabling features such as live map updates and real-time traffic information.
	On the other hand, the implementation of NS-COM assisted enhanced global navigation system also presents several challenges. There is a need for effective integration of navigation and communication systems, ensuring stability and reliability. NS-COM relies on significant spectrum resources, particularly for communication and positioning services in high-frequency bands, necessitating solutions to address spectrum scarcity issues. 
	
	\subsection{NS-COM Assisted Intelligent Unmanned System}
	
	The NS-COM technology, as shown in FSO communication is demonstrated in Fig. \ref{fig:5}, enables efficient collaborative operations between intelligent unmanned system, especially UAV swarms and unmanned underwater vehicle (UUV) swarms, extending beyond the boundaries of individual swarms. Specifically, NS-COM network can enable the coordination of autonomous swarms of unmanned systems, allowing them to operate in a synchronized manner. This could be particularly beneficial for tasks such as large-scale mapping, search and rescue operations, and environmental monitoring, where multiple unmanned systems can work together to achieve common goals more efficiently than a single unit.  Additionally, NS-COM technology enables multi-sensor data fusion, enhancing the system's environmental perception capabilities and improving its adaptability and robustness.
	Establishing stable and reliable communication links among UAV swarms, UUV swarms, and StratoSat is essential for data transmission and collaborative operations. However, adverse weather conditions, such as cloud cover, can affect the stability of communication links. Therefore, contingency plans for controlling communication link disruptions and strategies for link restoration after disruptions need to be devised. 
	To enhance the autonomous capabilities of unmanned systems, it is worth investigating to use the vast amounts of data collected by NS-COM network for machine learning model training.
	By analyzing patterns and learning from diverse scenarios, these systems can improve their decision-making processes, adapt to new environments, and perform complex tasks with minimal human intervention.
	\color{black}
	
	\subsection{FSO Communication}
	Given the substantial data exchanges between space, airborne and terrestrial network, the demand for high-speed connectivity escalates, making FSO communication a pivotal player in meeting these elevated data rate requisites.
	\color{black}
	FSO communication is demonstrated in Fig. \ref{fig:5}. Because of its highly focused beamwidth, FSO communication offers reduced or minimal interference, making it a promising solution of enhancing the QoS and spectral efficiency. The study in \cite{add2} explored the framework of combined FSO-RF transmission within a network integrating satellites and UAVs to address the growing demands of Internet of Remote Things devices, then maximized the  ergodic sum rate while adhering to constraints on total transmit power and QoS requirements for each Internet of Remote Things device.
	\color{black}
	FSO finds considerable utility in StratoSats, particularly in relaying scenarios, capitalizing on its strengths. Although the direct LoS transmission inherent in FSO signals can effectively prevent eavesdropping in contrast to RF signals, FSO technology also grapples with sensitivity to atmospheric obstacles like clouds and fog. Although these hindrances become trivial in StratoSat-satellite links, as StratoSats reside well above the cloud cover, the challenge yet persists when establishing optical communication between StratoSats and the ground due to the perpetual cloud impediment. 
	To maintain high data rate, photodiodes at the receiver that act as photodetectors need to be small. However, the laser beam widens in space, restricting the beam's travel range since the photodiode can only detect a smaller signal fraction over longer distances \cite{9611182}. The temperature and pressure of the atmosphere can also cause different effects on the performance of the FSO system, which is still unpredictable.
	\color{black}
	The performance of FSO communications is inevitably degraded due to limited transmit power affected by sky radiance and background shot noise, necessitating advanced transceiver design to improve system error performance. Moreover, present optoelectronic techniques predominantly depend on large front-end equipment, which remains inadequate for StratoSat applications due to restricted communication payload capacity. Hence, there is a need for a compact chip-based design for the FSO transceiver.
	\color{black}
	
	\section{Conclusion} 
	This article provided a thorough investigation of NS-COM that plays an indispensable role in SAGSIN. By analyzing the deployment, coverage, channel features across various network layers and the unique problems of NS-COM network, we showcased significance of NS-COM in enhancing the connectivity, coverage and transmission performance of the classical SAGSIN architecture. The examination of technical aspects, along with the exploration of applications and potential research directions, further reinforces the pivotal position of StratoSats in SAGSIN. We believe that the integration of NS-COM into SAGSIN will eventually revolutionize the communication paradigm on a global scale, paving the way for a future with ubiquitous broadband Internet access.
	
	\printbibliography
\end{document}